\documentclass[prd,twocolumn,showpacs,superscriptaddress,nofootinbib,floatfix,showkeys,10pt]{revtex4-2}

\usepackage{graphicx}
\usepackage{amsmath}
\usepackage{bm}
\usepackage{yhmath}
\usepackage{mathtools}
\usepackage{wasysym}
\usepackage[colorlinks,citecolor=blue,urlcolor=blue,linkcolor=blue]{hyperref}
\usepackage{subfigure}
\usepackage{color}
\usepackage{cases}
\usepackage{subfigure}
\usepackage{times}
\usepackage{dcolumn,booktabs,bm}
\usepackage{slashed}
\usepackage{amsfonts,amssymb,stmaryrd,latexsym,amsmath}
\usepackage{textcomp}
\usepackage{multirow}
\usepackage{cancel}
\usepackage{array}
\usepackage{orcidlink}
\usepackage{enumitem}
\usepackage{dashrule}

\definecolor{nicered}{rgb}{0.7,0.1,0.1}
\definecolor{nicegreen}{rgb}{0.1,0.5,0.1}
\definecolor{emph}{rgb}{1,0,0}
\definecolor{doub}{rgb}{0.7,0.2,1.0}
\definecolor{navyblue}{RGB}{0, 110, 184}

\hypersetup{colorlinks,citecolor=nicegreen,linkcolor=nicered,urlcolor=navyblue}

\newcommand{\clabel}[2][]{#2}

\renewcommand{\arraystretch}{1.8}

\begin{document}


\title{Trilepton and tetralepton bound and resonant states: the QED counterpart of multiquark states}

\author{Yao Ma\,\orcidlink{0000-0002-5868-1166}}
\affiliation{School of Physics and Center of High Energy Physics, Peking University, Beijing 100871, China}

\author{Lu Meng\,\orcidlink{0000-0001-9791-7138}}\email{lu.meng@rub.de}
\affiliation{Institut f\"ur Theoretische Physik II, Ruhr-Universit\"at Bochum,  D-44780 Bochum, Germany }

\author{Liang-Zhen Wen\,\orcidlink{0009-0006-8266-5840}}\email{
wenlzh\_hep-th@stu.pku.edu.cn}
\affiliation{School of Physics, Peking University, Beijing 100871, China}

\author{Shi-Lin Zhu\,\orcidlink{0000-0002-4055-6906}}\email{zhusl@pku.edu.cn}
\affiliation{School of Physics and Center of High Energy Physics, Peking University, Beijing 100871, China}

\begin{abstract}

This work presents the first prediction of tetralepton resonant states containing muons, extending beyond the simplest tetralepton system, dipositronium ($\mathrm{Ps}_2$). With the rapid advancements in experimental facilities, the production and study of these intriguing states may be  within reach. We perform a comprehensive analysis of S-wave trilepton and tetralepton systems within the framework of a QED Coulomb potential. We employ the Gaussian expansion method to solve the three- or four-body Schr\"odinger equation and utilize the complex scaling method to identify resonant states. We uncover a series of bound and resonant states in the trilepton systems $e^+e^+e^-$, $\mu^+\mu^+\mu^-$, $e^+e^+\mu^-$, and $\mu^+\mu^+e^-$, as well as the tetralepton systems $e^+e^+e^-e^-$, $\mu^+\mu^+\mu^-\mu^-$, and $\mu^+\mu^+e^-e^-$. The energies of these states range from $-30$ eV to $-1$ eV below the total mass of three or four leptons, with their widths varying from less than $0.01$ eV to approximately $0.07$ eV. Additionally, we calculate the spin configurations and root mean square radii of these states, providing insight into their spatial structures. No bound or resonant states are found in the trilepton $e^+\mu^+e^-$, $\mu^+e^+\mu^-$ systems, nor in the tetralepton $\mu^+e^+\mu^-e^-$ system. A comparison with fully heavy tetraquark systems reveals that the additional color degree of freedom in QCD results in the absence of low-energy bound and resonant states. However, this extra degree of freedom allows for a broader range of $J^{PC}$ quantum numbers to produce resonant states, highlighting the rich complexity of QCD systems.

\end{abstract}

\maketitle

\section{Introduction}~\label{sec:intro}

Few-lepton systems are composed of charged leptons, which primarily interact via electromagnetic forces, with their fundamental theory being quantum electrodynamics (QED). Due to their theoretical simplicity and precise calculability, leptonic systems held an important place in particle physics. The study of such systems dated back to 1946, when Wheeler first predicted the existence of polyelectrons~\cite{wheeler1946polyelectrons}. Subsequently, an increasing number of studies have explored few-electron bound states, such as the $e^+e^+e^-$~\cite{frost1964approximate,martin1992stability,Ho:1993zz,drake2002ground,khan2012hyperspherical,kylanpaa2012few,liverts2013three,deveikis2018two} and $e^{+} e^{+} e^{-} e^{-}$ ($\mathrm{Ps}_2$)~\cite{Hylleraas:1947zza,Ho:1986zz,ho1990positronium,Kinghorn:1993zz,kozlowski1993nonadiabatic,el1995variational,Bao:1998vy,Varga:1998ss,abdel1998existence,cheng1999existence,suzuki2000excited,bao2003lowest,bubin2006nonrelativistic,bubin2007relativistic,Puchalski:2008jj,hogaasen2010two,heyrovska2011atomic,rebane2012existence,teo2012exotic,matyus2012molecular,Varandas:2015ega,Assi:2023dlu,Munir:2023gsa} systems. At the same time, the exploration of multi-lepton bound states has expanded from electron systems to systems containing muons, including $\mu^+e^+e^-$~\cite{martin1992stability,khan2012hyperspherical,liverts2013three,frolov2017atomic,deveikis2018two}, $\mu^{+} \mu^{+} e^{-}$~\cite{frolov1999bound,liverts2013three}, $\mu^{+} e^{+} e^{-} e^{-}$~\cite{frolov1997positronium,bressanini1998stability,rebane2012existence,rebane2012symmetry,frolov2017atomic}, $\mu^{+} \mu^{+} \mu^{-} e^{-}$~\cite{rebane2012symmetry}, as well as systems with varying mass ratios, such as $X^{+} Y^{+} Y^{-} Y^{-}$ (where $X$ and $Y$ denote particles with different masses)~\cite{richard1994stability,varga1997stability,bressanini1998stability,mitroy2004stability,rebane2012existence,rebane2012symmetry}, $X^{+} X^{+} Y^{-} Y^{-}$~\cite{richard1994stability,varga1997stability,bressanini1998stability,rebane2012existence,rebane2012stability,rebane2012symmetry}, $X^{+} X^{-} Y^{+} Y^{-}$~\cite{Bressanini:1997zz,abdel1998existence,rebane2001stability,rebane2003binding,gridnev2005proof,VanHooydonk:2004su,el2007investigation,rebane2012existence,rebane2012symmetry} systems, see reviews~\cite{Suzuki:1998wut,richard2002stability,armour2005stability,emami2015short,emami2021review}. These exotic systems are not only of interest to particle physicists but also to researchers in atomic and molecular physics.

However, researches on the few-lepton systems have primarily focused on bound states, the exploration of resonant states remains limited. Some progress has been made, such as the investigation of resonant states in the $e^+e^+e^-$~\cite{Ho:1979zz,ho1984doubly,rost1992positronium,ivanov1999high,Usukura:2002zz,suzuki2004stochastic,kar2009d,ho2012complex,matyus2013calculation,liverts2013three,igarashi2016broad,Kar:2018sax,Kar:2018vmc,kar2019calculations} and $e^+e^+e^-e^-$~\cite{ho1989resonant,Usukura:2002zz,bao2003lowest,suzuki2004stochastic,dirienzi2010resonances,matyus2013calculation,zhang2020doubly} systems. For systems with muons, only the $e^+e^+\mu^-$ resonant states have been investigated~\cite{Ho:1979zz,liverts2013three}. Up to now, no studies have yet explored the tetralepton resonance states involving muons. The limitation is partly due to the significantly increased complexity of four-body systems compared to three-body systems. In fact, calculating resonance states requires far more computational resources than solving bound states, which was difficult to achieve in the past. Furthermore, for systems containing both muons and electrons, the significant mass difference between the muon and the electron makes the calculations even more challenging. Although the mass difference in hydrogen molecules is even greater, previous studies on hydrogen molecules have mostly relied on the Born-Oppenheimer approximation to reduce the problem to an effective two-body system~\cite{heitler1927wechselwirkung}, 
and no resonance states have been explored in such system. As a result, the physical characteristics of tetralepton resonance states involving muons have yet to be revealed clearly. However, with advancements in computational techniques, we are now able to partially overcome these challenges. In this work, we present the first prediction of tetralepton resonant states beyond the simplest 4-body system, $\mathrm{Ps}_2$, opening up a new type of state. This paves the way for a more comprehensive exploration of few-lepton systems, providing new opportunities to uncover their physical characteristics.

Experimental efforts in the search for few-lepton states have already yielded significant results. The first experiment to prove the existence of the positronium negative ion was performed by Mills in 1981~\cite{mills1981observation}, demonstrating the feasibility of detecting such exotic tetralepton states. The existence of the $\mathrm{Ps}_2$ molecule bound state was experimentally seen by Cassidy and Mills in 2007~\cite{cassidy2007production}. With advancements in experimental facilities, the production of tetralepton states containing muons may be possible. For example, the Super $\tau$-Charm Facility~\cite{Achasov:2023gey} may provide an potential platform for producing and studying these tetralepton states. Additionally, with the upgrade of muon beams, such states could potentially be produced by inserting a $\mu$ beam into electron gas~\cite{Doble:1994np,Prokscha:2008zz,Ganguly:2022ufq,Kanda:2023gqp}.

\begin{figure}[htbp]
  \centering
  \includegraphics[width=0.4\textwidth]{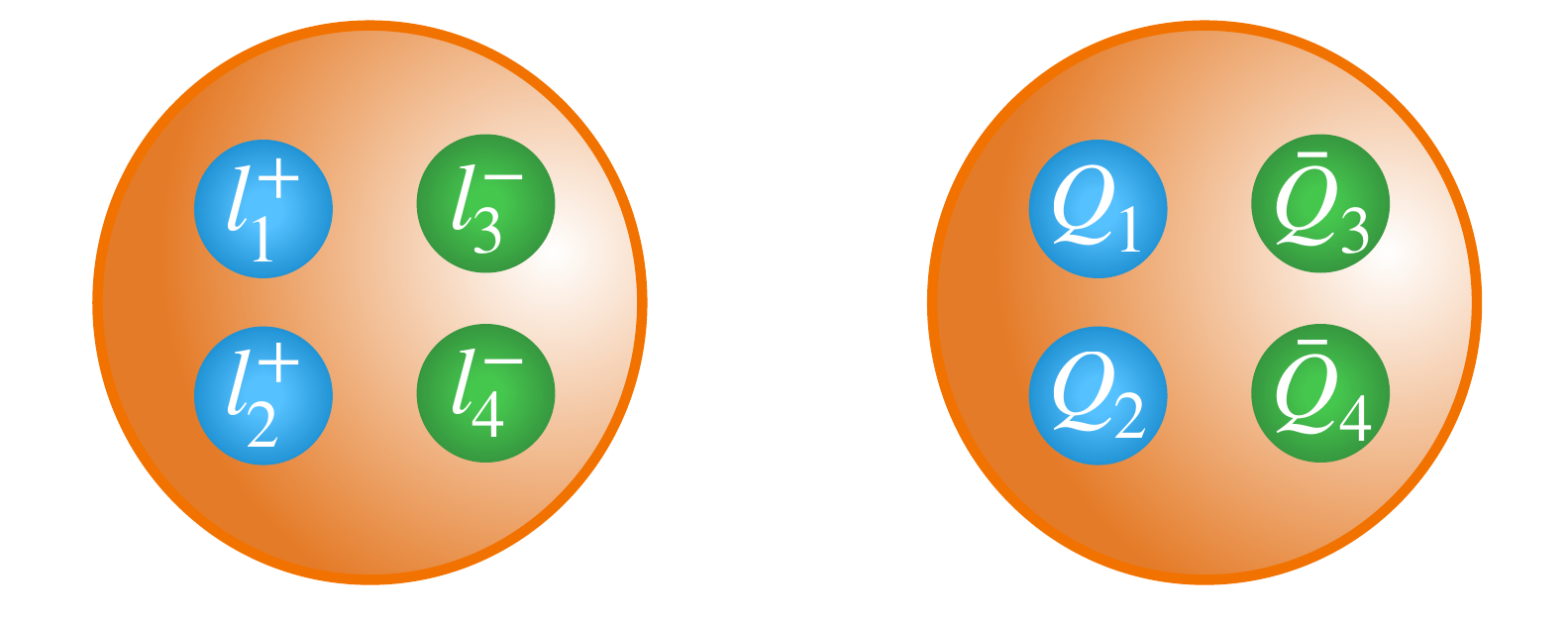} 
  \caption{\label{fig:ppmm} Tetralepton system and tetraquark system. }
    \setlength{\belowdisplayskip}{1pt}
\end{figure}

In addition to the intrinsic value of exploring tetralepton states themselves, another reason why the tetralepton system is particularly significant is that it shares many similarities with the recently popular tetraquark states (see ~\cite{Chen:2016qju,Esposito:2016noz,Guo:2017jvc,Liu:2019zoy,Brambilla:2019esw,Chen:2022asf,Meng:2022ozq} for reviews), which have been intensively studied over the past two decades since the discovery of $X(3872)$ in 2003~\cite{Belle:2003nnu}. They both contain two particles and two antiparticles as shown in Fig.~\ref{fig:ppmm}. 
For heavy tetraquark systems, the small quark kinetic energy leads to a shorter distance between the quarks, thereby making the color-electric Coulomb interaction dominant, just as in the tetralepton system. Therefore, the tetralepton system can be regarded as the QED counterpart of the tetraquark state.

Studying such tetralepton resonant states could provide deeper insights into the nature of tetraquark states. Notably, all experimentally observed fully heavy tetraquark candidates, such as $X(6900)$ and $X(7200)$~\cite{LHCb:2020bwg,CMS:2023owd,ATLAS:2023bft},  lie above the di-charmonium thresholds. This suggests that they are resonant states with finite lifetimes. As a heavy quark system dominated by Coulomb interactions, it should exhibit properties similar to those of the tetralepton system. Meanwhile, in the doubly heavy tetraquark system, although the experimentally discovered $T_{cc}^+(3875)$ is a shallow bound state~\cite{LHCb:2021auc}, numerous theoretical studies have predicted the existence of resonance states in such systems~\cite{Richard:2022fdc,Yang:2019itm,Albaladejo:2021vln,Meng:2023for,Meng:2024yhu,Wu:2024zbx}.
Therefore, even for the purpose of comparison with tetraquark states, exploring tetralepton resonant states above the threshold is highly worthwhile. 

In this work, we investigate the bound and resonant states in trilepton systems $e^+e^+e^-$, $\mu^+\mu^+\mu^-$, $e^+e^+\mu^-$, $\mu^+\mu^+e^-$, $e^+\mu^+e^-$, $\mu^+e^+\mu^-$ and tetralepton systems $e^+e^+e^-e^-$, $\mu^+\mu^+\mu^-\mu^-$, $\mu^+\mu^+e^-e^-$, $\mu^+e^+\mu^-e^-$. We utilize the
complex scaling method~\cite{Aguilar:1971ve,Balslev:1971vb,aoyama2006complex} to obtain possible bound states and resonant states simultaneously. We employ the
Gaussian expansion method~\cite{Hiyama:2003cu} to solve the $n$-body Schr\"odinger equation, which has been successfully used in
our previous work on bound states~\cite{Meng:2023jqk,Zhu:2024hgm} and resonant states~\cite{Chen:2023syh,Wu:2024euj,Wu:2024hrv,Ma:2024vsi,Wu:2024zbx}. We show the spectra, spatial structures and spin configurations of the obtained states.

This paper is arranged as follows. 
In Sec.~\ref{sec:framework}, we introduce the theoretical framework, including the Hamiltonian with the Coulomb potential, the construction of the wave function, the complex scaling method, and the approach to analyzing the spatial structures.
In Sec.~\ref{sec:results}, we present the numerical results for the properties of the trilepton and tetralepton bound and resonant states.
Finally, we give a brief summary and discussion in Sec.~\ref{sec:sum}.

\section{Theoretical framework}~\label{sec:framework}

\subsection{Hamiltonian}~\label{subsec:Hamiltonian}

The nonrelativistic Hamiltonian of an $n$-body system reads
\begin{align}\label{eq:Hamiltonian}
H=\sum_i^n\left(m_i+\frac{\boldsymbol{p}_i^2}{2m_i}\right)+\sum_{i<j=1}^n V_{i j}\,,
\end{align}
where $m_i$ and $\boldsymbol{p}_i$ are the mass and momentum of particle $i$. \clabel[relativisticEffects]{Since the order of $p/m \sim 10^{-3}$ in QED systems, the relativistic effect are neglected within the precision of our calculations. }

We use the Coulomb potential for leptonic systems
\begin{align}\label{eq:Coulomb}
    V_{i j}(r)&=\frac{Q_i Q_j}{r_{i j}},
\end{align}
where $Q_i$ represents the charge of lepton $i$. 
\clabel[spinEffects]{ We neglect the spin-spin interaction effect because it is suppressed by a factor of $1/m^2$, and the binding energy in QED systems is much smaller than the lepton mass, making this suppression extremely significant.}
The lepton masses and the fine-structure constant $\alpha$ are taken from Particle Data Group~\cite{ParticleDataGroup:2024cfk}. The calculated masses of the two-lepton bound states are presented in Table~\ref{tab:2body}. Our results match the analytical solutions exactly. Since no spin-dependent interaction is introduced, their spin and C-parity are degenerate.

\begin{table}[htbp] 
\renewcommand{\arraystretch}{1.4}
\centering
\caption{\label{tab:2body} The exact binding energies $\Delta E_{\mathrm{exact}}$, calculated binding energies $\Delta E_{\mathrm{calc}}$ and rms radii $r^{\mathrm{rms}}_{\mathrm{calc}}$ of the $l^+l^{(')-}$ systems. Ps represents the positronium $e^+e^-$.}
\begin{tabular*}{\hsize}{@{}@{\extracolsep{\fill}}lccccc@{}}

\hline\hline
$J^{PC}$ & System & $\Delta E_{\mathrm{exact}}$  & $\Delta E_{\mathrm{calc}}$  & $r^{\mathrm{rms}}_{\mathrm{exact}}$& $r^{\mathrm{rms}}_{\mathrm{calc}}$\\
\hline 
$0^{-+} / 1^{--}$ & $\mathrm{Ps}(1 S)$ & -6.80 eV  & -6.80 eV & 0.18 nm & 0.18 nm\\
& $\mathrm{Ps}(2 S)$ & -1.70 eV & -1.70 eV & 0.69 nm & 0.69 nm\\
& $\mathrm{Ps}(3 S)$ & -0.76 eV & -0.76 eV & 1.52 nm & 1.52 nm\\
\cline{2-6} 
& $\mu^{+} e^{-}(1 S)$ & -13.5 eV & -13.5 eV & 0.09 nm & 0.09 nm \\
& $\mu^{+} e^{-}(2 S)$ & -3.4 eV & -3.4 eV & 0.34 nm & 0.34 nm \\
& $\mu^{+} e^{-}(3 S)$ & -1.5 eV & -1.5 eV & 0.76 nm & 0.76 nm\\
\cline{2-6} 
& $\mu^{+} \mu^{-}(1 S)$ & -1.41 keV & -1.41 keV & 0.9 pm & 0.9 pm \\
& $\mu^{+} \mu^{-}(2 S)$ & -0.35 keV & -0.35 keV & 3.3 pm & 3.3 pm\\
& $\mu^{+} \mu^{-}(3 S)$ & -0.16 keV & -0.16 keV & 7.4 pm & 7.4 pm\\

\hline \hline
\end{tabular*}
\end{table}

\subsection{Wave function construction}~\label{subsec:wavefunction}

The wave function bases of the $l_1^+l_2^+l_3^-$ trilepton and $l_1^+l_2^+l_3^-l_4^-$ tetralepton systems can be expressed as the direct product of spatial wave function $\phi$ and spin wave function $\chi_{s}$.
\begin{equation}\label{eq:Abasis}
\psi=\mathcal{A}\left(\phi \otimes \chi_{s}\right),
\end{equation}
where $\mathcal{A}$ is the antisymmetrization operator, representing the exchange of identical leptons. For instance, for the $e^+e^+e^-e^-$ and  $\mu^+\mu^+e^-e^-$ system, $\mathcal{A}=\left(1-P_{12}\right)\left(1-P_{34}\right)$, where $P_{i j}$ permutes the $i$th and $j$th (anti)leptons.

For the spatial wave function, the Gaussian expansion method (GEM)~\cite{Hiyama:2003cu} is employed. Namely, the spatial wave function is expanded using the following basis:
\begin{equation}\label{eq:basisSpace}
\phi_{n l m}(\boldsymbol{r})=\sqrt{\frac{2^{l+5 / 2}}{\Gamma\left(l+\frac{3}{2}\right) r_n^3}}\left(\frac{r}{r_n}\right)^l e^{-\frac{r^2}{r_n^2}} Y_{l m}(\hat{r}),
\end{equation}
where the $r_n$ is taken in geometric progression, $r_n=r_0 a^n$. $Y_{l m}$ is the spherical harmonics.

\begin{figure}[htbp]
  \centering
  \includegraphics[width=0.48\textwidth]{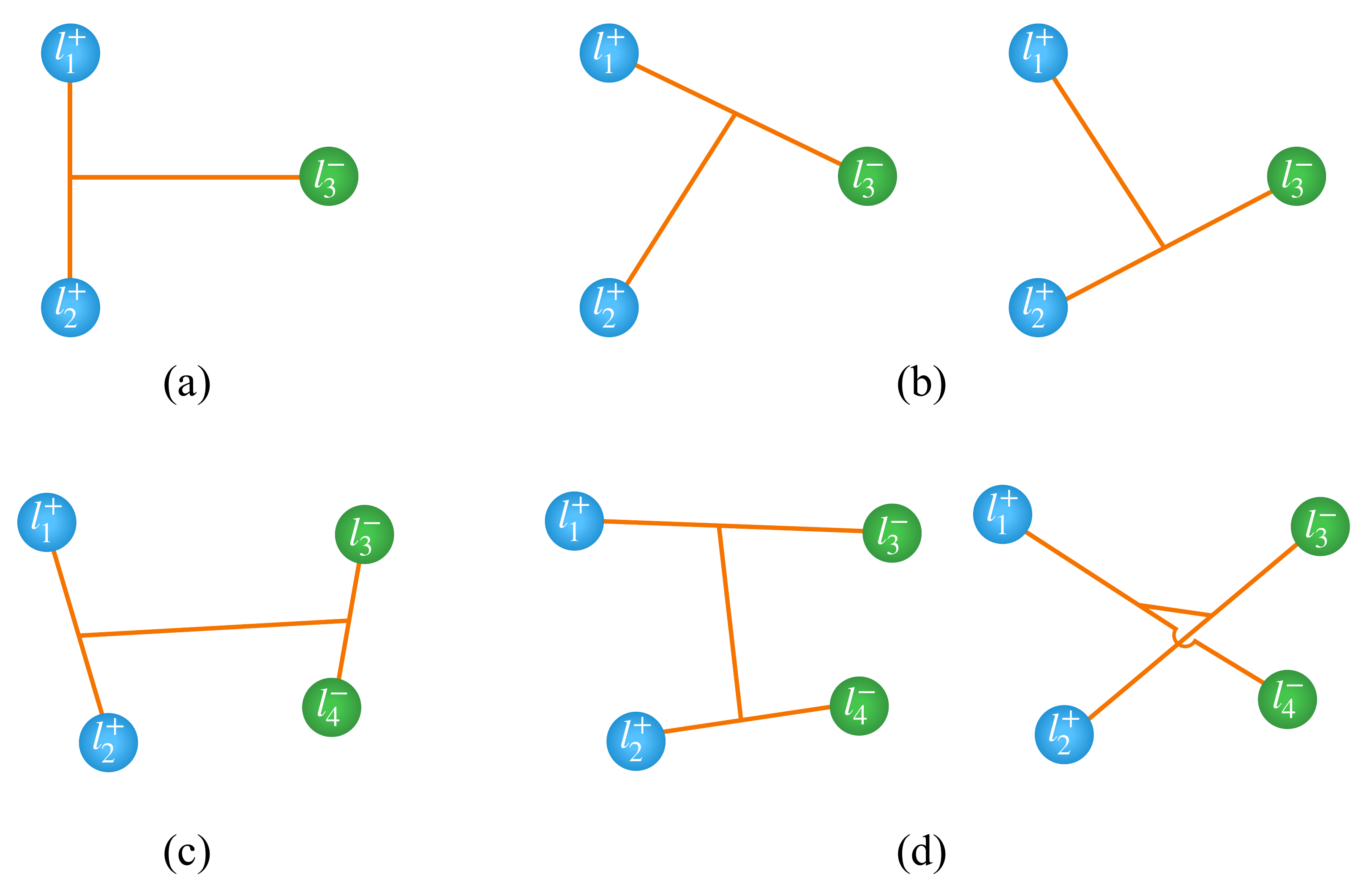} 
  \caption{\label{fig:structure_34l} (a) and (b): two types of structures for the trilepton system. (c) and (d): two types of structures for the tetralepton system. }
    \setlength{\belowdisplayskip}{1pt}
\end{figure}

For an $n$-body system, one can work on the Jacobi coordinates to exclude the center-of-mass motion. In principle, one could construct a complete basis functions using any type of Jacobi coordinates. However, to get rid of the complexity from the angular momentum, we use only S-wave bases constructed in different types of Jacobi coordinates to partially compensate for higher partial waves (see Ref.~\cite{Meng:2023jqk} for details). In this work, we include different types of Jacobi coordinates for the trilepton and tetralepton systems as shown in Fig.~\ref{fig:structure_34l}. Their Jacobi coordinates can be expressed as:
\begin{align}
\boldsymbol{r}_{i j} & =\boldsymbol{r}_i-\boldsymbol{r}_j ,\\
\boldsymbol{r}_{ij,k} & =\frac{m_i \boldsymbol{r}_i+m_j \boldsymbol{r}_j}{m_i+m_j}-\boldsymbol{r}_k ,\\
\boldsymbol{r}_{ij,kl} & =\frac{m_i \boldsymbol{r}_i+m_j \boldsymbol{r}_j}{m_i+m_j}-\frac{m_k \boldsymbol{r}_k+m_l \boldsymbol{r}_l}{m_k+m_l} .
\end{align}
The settings of the basis parameters for each system in our calculation are summarized in Appendix~\ref{app:Gaussian_par}.

For the spin wave function, the bases for total spin $S$ are

\begin{itemize}
    \item $\mathrm{3-body}:$
    \begin{equation}
    \begin{aligned}
    & S=\frac{1}{2}:\left\{
    \begin{array}{l}
    {\left[\left(l_1^{+} l_2^{+}\right)_{0} l_3^{-}\right]_{\frac{1}{2}}} \\
    {\left[\left(l_1^{+} l_2^{+}\right)_{1} l_3^{-}\right]_{\frac{1}{2}}}
    \end{array}\right., \\
    & S=\frac{3}{2}:
    \begin{array}{l}
    {\left[\left(l_1^{+} l_2^{+}\right)_{1} l_3^{-}\right]_{\frac{3}{2}}}
    \end{array}.
    \end{aligned}
    \end{equation}
\end{itemize}

\begin{itemize}
    \item $\mathrm{4-body}:$
    \begin{equation}
    \begin{aligned}
    & S=0:\left\{\begin{array}{l}
    {\left[\left(l_1^{+} l_2^{+}\right)_0\left(l_3^{-} l_4^{-}\right)_0\right]_0} \\
    {\left[\left(l_1^{+} l_2^{+}\right)_1\left(l_3^{-} l_4^{-}\right)_1\right]_0}
    \end{array},\right. \\
    & S=1:\left\{\begin{array}{l}
    {\left[\left(l_1^{+} l_2^{+}\right)_0\left(l_3^{-} l_4^{-}\right)_1\right]_1} \\
    {\left[\left(l_1^{+} l_2^{+}\right)_1\left(l_3^{-} l_4^{-}\right)_0\right]_1} \\
    {\left[\left(l_1^{+} l_2^{+}\right)_1\left(l_3^{-} l_4^{-}\right)_1\right]_1}
    \end{array},\right. \\
    & S=2:\begin{array}{l}
    {\ \ \ \left[\left(l_1^{+} l_2^{+}\right)_1\left(l_3^{-} l_4^{-}\right)_1\right]_2}
    \end{array}.
    \end{aligned}
    \end{equation}
\end{itemize}
Since the potential is spin-independent, there is no coupling between these spin channels.

Besides the antisymmetrization operation in Eq. (\ref{eq:Abasis}), since the $l^+ l^+ l^- l^-$ system carries definite C-parity, it is necessary to add (or subtract) the C-transformation of each term in Eq. (\ref{eq:Abasis}) to ensure that the overall wave function has the desired positive (or negative) C-parity. The specific C-transformation is
\begin{align}
&\left[\left(l_1^{+} l_2^{\prime+}\right)^{s_1}\left(l_3^{-} l_4^{\prime-}\right)^{s_2}\right]^S\phi\left(\boldsymbol{r}_1, \boldsymbol{r}_2, \boldsymbol{r}_3, \boldsymbol{r}_4\right)\nonumber\\
\xrightarrow{C}&(-1)^{S-s_1-s_2}\left[\left(l_1^{+} l_2^{\prime+}\right)^{s_2}\left(l_3^{-} l_4^{\prime-}\right)^{s_1}\right]^S\phi(\boldsymbol{r}_{3},\boldsymbol{r}_{4},\boldsymbol{r}_{1},\boldsymbol{r}_{2})\nonumber.
\end{align}
The particle $l$ and $l'$ can be different leptons.

\subsection{Complex scaling method}~\label{subsec:method}

The complex scaling method (CSM) is a direct approach to obtain the energies and the decay widths of resonant states in a many-body system by performing an analytical continuation of the Schr\"odinger equation~\cite{Aguilar:1971ve,Balslev:1971vb,aoyama2006complex}. This is achieved by carrying out a complex rotation on the coordinate $\boldsymbol{r}$ and momentum $\boldsymbol{p}$, given by
\begin{align}\label{eq:complexRotation}
U(\theta) \boldsymbol{r}=\boldsymbol{r} e^{i \theta}, \quad U(\theta) \boldsymbol{p}=\boldsymbol{p} e^{-i \theta}.
\end{align}
Under the rotation, the Hamiltonian in Eq. (\ref{eq:Hamiltonian}) becomes
\begin{equation}\label{eq:HamiltonianComplex}
H(\theta)=\sum_{i=1}^n\left(m_i+\frac{p_i^2 e^{-2 i \theta}}{2 m_i}\right)+\sum_{i<j=1}^n V_{i j}\left(r_{i j} e^{i \theta}\right).
\end{equation}
Meanwhile, for the resonant states with pole positions within the range of the rotated angle, their wave functions become normalizable by integration, thereby solvable through localized Gaussian bases in the same way as bound states. As a result, solving the complex-scaled Schr\"odinger equation will simultaneously yield the eigenenergies of bound states and resonant states within the rotated angle.

\begin{figure}[htbp]
  \centering
  \includegraphics[width=0.45\textwidth]{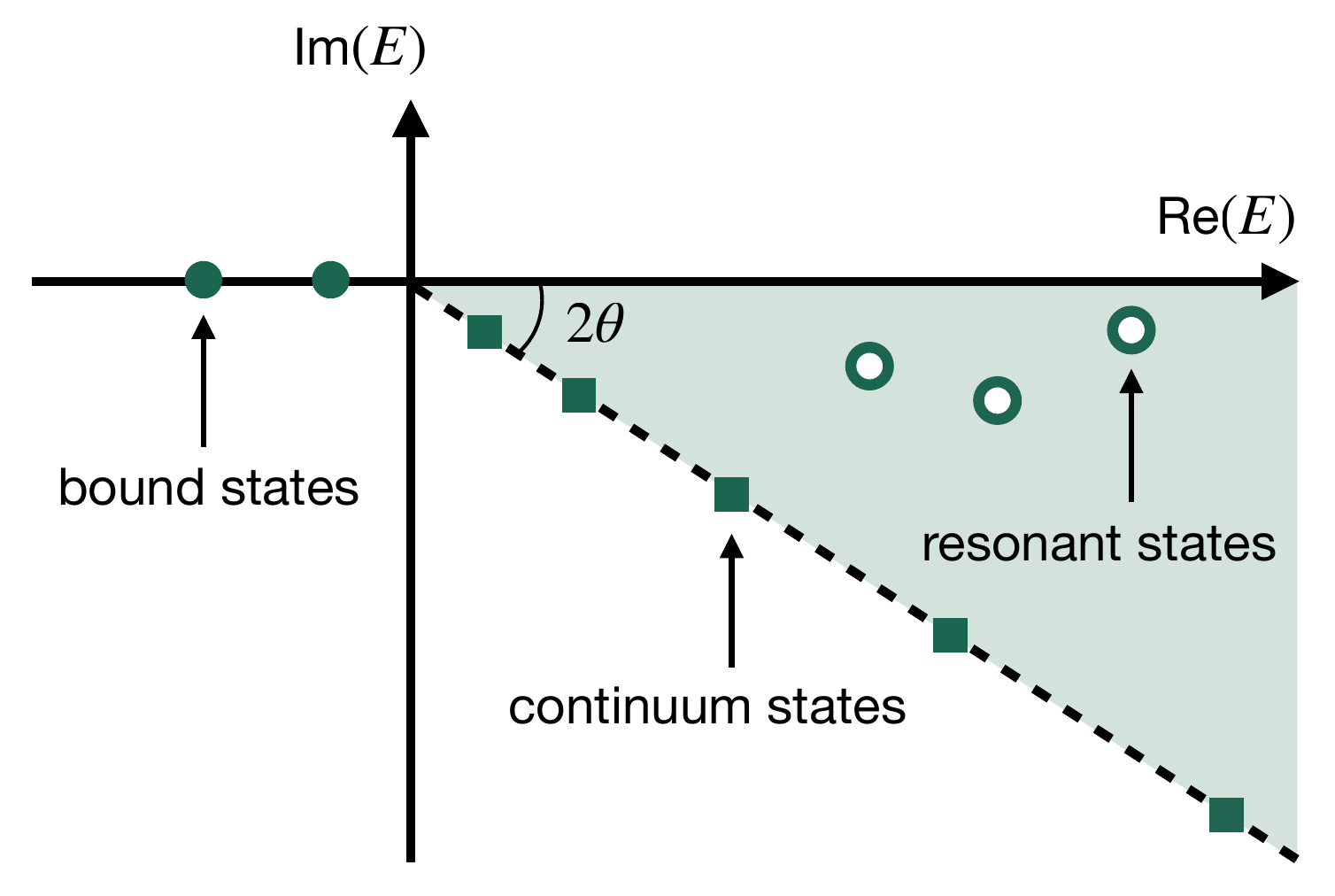} 
  \caption{\label{fig:CSMpattern}Typical eigenvalue distribution of the complex scaled $H(\theta)$ for two-body systems. The solid circles on the negative horizontal axis represent bound states, and the open circles denote resonant states. The continuum states (squares) align along the $2 \theta$-line (dashed line).}
    \setlength{\belowdisplayskip}{1pt}
\end{figure}

A typical pattern of the solved eigenenergies in the complex energy plane is shown in Fig.~\ref{fig:CSMpattern}. The bound states lie on the negative real axis of the energy plane. The continuum states align along beams originating from thresholds with $\operatorname{Arg}(E)=-2 \theta$. The resonant states with mass $M_R$ and width $\Gamma_R$ are located at $E_R=M_R-i \Gamma_R / 2$, and only those within $\left|\operatorname{Arg}\left(E_R\right)\right|<2 \theta$ can be solved. The positions of the bound and resonant states remain unchanged with the variation of the rotation angle. One can find more details in Refs.~\cite{Lin:2022wmj,Chen:2023eri,Chen:2023syh}.

\clabel[analyticStructure]{Actually, the analytic structure of the S-matrix for three-body and four-body systems are more complicated than two-body systems~\cite{Sadasivan:2021emk,Dawid:2023kxu}.   While we have not fully explored this complexity, the CSM remains effective in identifying poles near the physical region, just as it does for resonances above the threshold in two-body systems.}

\subsection{Spatial structure}

The root-mean-square (rms) radius is a good physical quantity for reflecting the spatial structure of the states. The definition of the rms radius under CSM is 
\begin{equation}
r_{i j}^{\mathrm{rms,C}} \equiv \operatorname{Re}\left[\sqrt{\frac{\left(\Psi(\theta)\left|r_{i j}^2 e^{2 i \theta}\right| \Psi(\theta)\right)}{\left(\Psi(\theta) \mid \Psi(\theta)\right)}}\right],
\end{equation}
where the $\Psi(\theta)$ is the obtained complex wave function of the $n$-lepton state. The round bra-ket represents the so-called c-product~\cite{Romo:1968tcz} defined as
\begin{equation}
\left(\phi_n \mid \phi_m\right) \equiv \int \phi_n(\boldsymbol{r}) \phi_m(\boldsymbol{r}) \mathrm{d}^3\boldsymbol{r},
\end{equation}
without taking complex conjugate of the bra-state. This procedure ensures the function inside the integral is analytic, thereby the expectation value of the physical quantity remains stable as the rotation angle changes. The rms radius calculated from the c-product is generally not real; however, its real part can still reflect the internal lepton clustering behavior if the resonant state is not too broad, as discussed in Ref.~\cite{homma1997matrix}.

\section{Numerical results}~\label{sec:results}

\subsection{Trilepton systems}\label{subsec:3l}

\subsubsection{$e^+e^+e^-$ and $\mu^+\mu^+\mu^-$}\label{subsubsec:eee}

\begin{figure*}[htbp]
  \centering
  \includegraphics[width=0.8\textwidth]{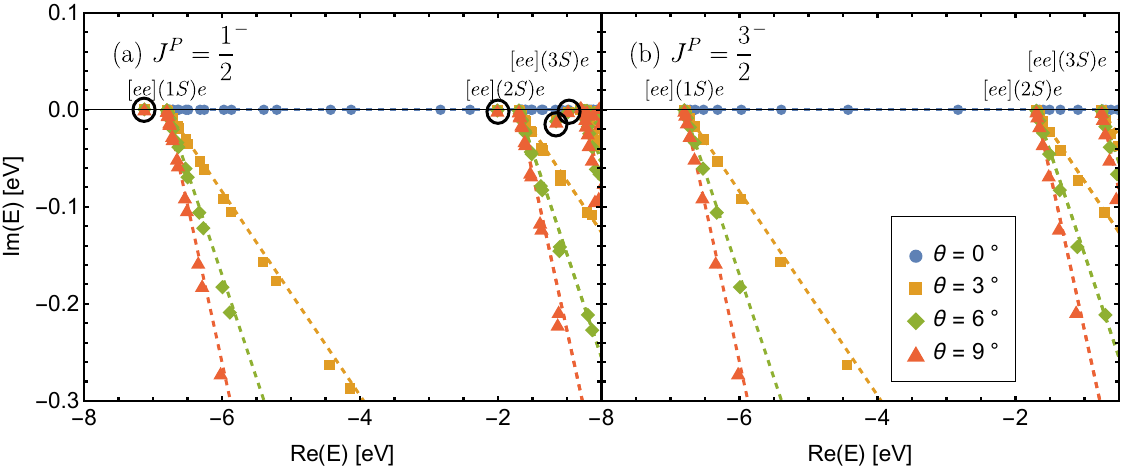} 
  \caption{\label{fig:eee} The complex energy eigenvalues of the $e^+e^+e^-$ states with varying $\theta$ in the CSM. The solid lines represent the continuum lines rotating along $\mathrm{Arg}(E)=-2\theta$. The resonant states do not shift as $\theta$ changes and are marked out by the black circles. }
    \setlength{\belowdisplayskip}{1pt}
\end{figure*}

The complex eigenenergies yielded for $e^+e^+e^-$ system with $J^{P}=\frac{1}{2}^{-},\frac{3}{2}^{-}$ are shown in Fig.~\ref{fig:eee}. We choose different complex scaling angles $\theta=3^\circ,6^\circ,9^\circ$ to distinguish bound states and resonant states, which remain stationary as the angle changes, from the continuum states. The markers that align along beams originating from positronium-positron thresholds with $\operatorname{Arg}(E) = -2 \theta$ are the continuum state eigenenergies. The bound and resonant states are marked out by black circles. A bound state is found below the lowest $[ee](1S)e$ threshold and a series of resonant states are obtained. Their complex energies, spin configurations, and rms radii are summarized in Table~\ref{tab:3body}.

\begin{table*}[htbp] 
\renewcommand{\arraystretch}{1.4}
\centering
\caption{\label{tab:3body} The complex energies $\Delta E-i \Gamma / 2$ (in eV), spin configuration and rms radii (in nm) of the $l_1^+l_2^+l_3^-=e^+e^+e^-$, $e^+e^+\mu^-$ and $\mu^+\mu^+e^-$ bound and resonant states with $J^P=\frac{1}{2}^-$. $\Delta E$ represents the binding energy relative to the three body threshold. In the ``type" column, ``B" represents bound states, while ``R" denotes resonant states. The spin configuration is represented as $[s_{12},s_{3}]_S$.  In the rightmost column, results from the literature using the variational method with different wave functions are presented for comparison. These include the Pekeris method~\cite{frost1964approximate,liverts2013three}, the Hylleraas-type wave function\cite{Ho:1979zz,Ho:1993zz}, the stochastic variational approach with correlated Gaussians~\cite{suzuki2004stochastic}, and the exponential variational expansion with nonlinear parameters~\cite{frolov1999bound,frolov2017atomic}. All resonance results were obtained using the complex scaling method.
 }
\begin{tabular*}{\hsize}{@{}@{\extracolsep{\fill}}lcccccc@{}}

\hline\hline
system & $\Delta E-i \Gamma / 2$ (This work) & type & configuration &$r^{\mathrm{rms}}_{l^+l^+}$ &$r^{\mathrm{rms}}_{l^+l^{(')-}}$&$\Delta E-i \Gamma / 2$ \\

\hline 
$e^+e^+e^-$ & $-7.12$ & B & $[0,\frac{1}{2}]_{\frac{1}{2}}$ & $0.51$ & $0.37$ & $-7.13$~\cite{frost1964approximate,Ho:1993zz} \\
& $-2.01$ & R & $[0,\frac{1}{2}]_{\frac{1}{2}}$ & $1.32$ & $0.83$  & $-2.07-0.0006i$~\cite{Ho:1979zz,suzuki2004stochastic,liverts2013three}\\
& $-1.16-0.01i$ & R & $[0,\frac{1}{2}]_{\frac{1}{2}}$ & $1.85$ & $1.31$ &...  \\
& $-0.98$ & R & $[0,\frac{1}{2}]_{\frac{1}{2}}$ & $2.62$ & $1.61$  & $-0.96-0.001i$~\cite{Ho:1979zz,suzuki2004stochastic}\\
\hline 
$e^+e^+\mu^-$ & $-14.29$ & B & $[0,\frac{1}{2}]_{\frac{1}{2}}$ & $0.27$ & $0.18$ & $-14.29$~\cite{liverts2013three,frolov2017atomic}\\
& $-4.03-0.02i$ & R & $[0,\frac{1}{2}]_{\frac{1}{2}}$ & $0.75$  & $0.44$  & $-4.03-0.02i$~\cite{liverts2013three} \\
& $-1.86-0.02i$ & R & $[0,\frac{1}{2}]_{\frac{1}{2}}$ & $1.64$ & $0.93$   &... \\
\hline 
$\mu^+\mu^+e^-$  & $-15.61$ & B & $[0,\frac{1}{2}]_{\frac{1}{2}}$ &$0.12$  & $0.11$ & $-15.92$~\cite{frolov1999bound,liverts2013three}\\
 & $-14.92$ & B & $[0,\frac{1}{2}]_{\frac{1}{2}}$ & $0.15$  & $0.13$  & $-15.21$~\cite{liverts2013three}\\
 & $-14.37$ & B & $[0,\frac{1}{2}]_{\frac{1}{2}}$ & $0.18$ & $0.15$  & $-14.62$~\cite{liverts2013three} \\
 & $-13.95$ & B & $[0,\frac{1}{2}]_{\frac{1}{2}}$ & $0.22$ & $0.18$ &... \\
 & $-13.67$ & B & $[0,\frac{1}{2}]_{\frac{1}{2}}$ & $0.30$  & $0.23$ &... \\
 & $-13.54$ & B & $[0,\frac{1}{2}]_{\frac{1}{2}}$ & $0.56$  & $0.41$ &... \\
 & $-3.63$ & R & $[0,\frac{1}{2}]_{\frac{1}{2}}$ & $0.59$  & $0.51$  &... \\
 & $-3.55$ & R & $[0,\frac{1}{2}]_{\frac{1}{2}}$ & $0.67$ & $0.56$  &... \\
 & $-3.48$ & R & $[0,\frac{1}{2}]_{\frac{1}{2}}$ & $0.76$ & $0.62$  &... \\

\hline \hline
\end{tabular*}
\end{table*}

The binding energy of the bound state is consistent with previous studies~\cite{frost1964approximate,martin1992stability,Ho:1993zz}. And the energies and widths of resonant states are consistent with Ref.~\cite{Ho:1979zz,suzuki2004stochastic,liverts2013three}. This consistency confirms the accuracy and reliability of our calculations.

\clabel[annihilation]{We do not consider the annihilation of positron and electron here because the contribution of this effect to the width is negligible compared to the precision of our calculations. To illustrate this point, we start with the two-body positronium system. The annihilation decay rates of ortho- and para-positronium are approximately $10^6\mathrm{~s}^{-1}\left(\sim 10^{-9} \mathrm{eV}\right)$  and $10^9\mathrm{~s}^{-1}\left(\sim 10^{-6} \mathrm{eV}\right)$, respectively~\cite{Karshenboim:2003vs}. In multilepton systems, the dominant annihilation decay channel is the annihilation of a pair of leptons into $2 \gamma$. For example, $\Gamma_{2 \gamma}\left(\mathrm{Ps}^{-}\right) \sim 10^9 \mathrm{~s}^{-1}\left(\sim 10^{-6} \mathrm{eV}\right)$~\cite{Frolov:2009qi}. This width is four orders smaller than the rearrangement fall-apart widths of $10^{13} \mathrm{~s}^{-1}\left(\sim 10^{-2} \mathrm{~eV}\right)$ in our work and is therefore negligible.}


For convenience, we label $e^+e^+e^-$ sequentially as 1, 2, and 3. The results show that only the total spin $S=\frac{1}{2}$ system can form bound and resonant states. And from the spin configuration, we find that all these states are in $\left[s_{12}, s_3\right]_S=\left[0, \frac{1}{2}\right]_{\frac{1}{2}}$ component. That is, the two identical $e^+$ have their spins anti-aligned. The absence of spin mixing is due to the spin-independent potential. This result is reasonable because spin anti-alignment is more likely to be bound. Due to the constraint of the Pauli principle, spin anti-alignment corresponds to a symmetric spatial wavefunction, which allows for an $S$-wave configuration. In contrast, spin alignment leads to an antisymmetric spatial wavefunction, requiring higher partial waves, and thus results in higher energy levels. 

From the rms radii of the $e^+e^+e^-$ states shown in Table~\ref{tab:3body}, we can see that the distance between $e^+$ and $e^+$ is slightly smaller than twice the distance between $e^+$ and $e^-$. This indicates that the electron is shared by the two positrons, and the state exhibits a structure similar to a covalent one-electron bond. The state below the second threshold can be viewed as the radial excitation of the bound state below the first threshold. They have the same component and similar structures.

\begin{figure*}[htbp]
  \centering
  \includegraphics[width=0.8\textwidth]{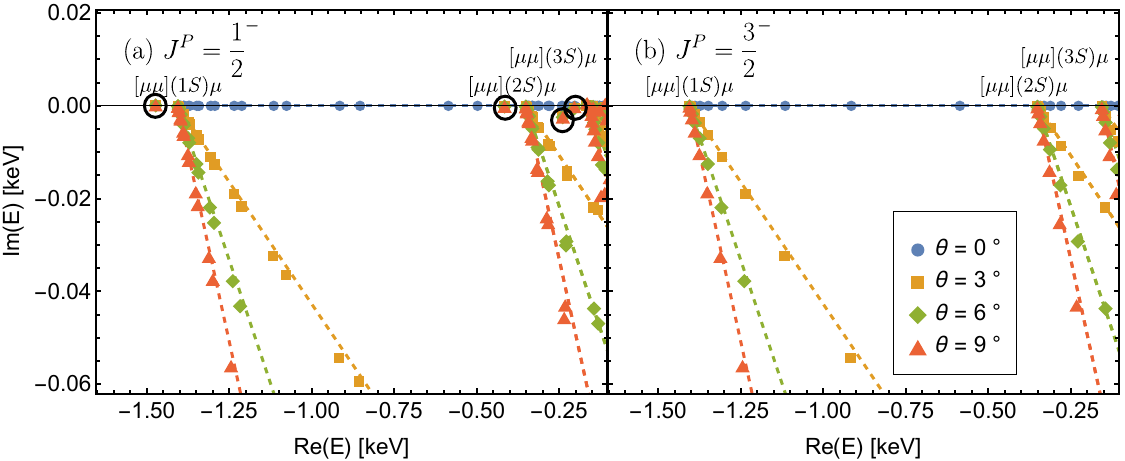} 
  \caption{\label{fig:mumumu}The complex energy eigenvalues of the $\mu^+\mu^+\mu^-$ states with varying $\theta$ in the CSM.}
    \setlength{\belowdisplayskip}{1pt}
\end{figure*}

The results of $\mu^+\mu^+\mu^-$ system are shown in Fig.~\ref{fig:mumumu} and Table~\ref{tab:3body_mumumu}. The pattern of the $\mu^+\mu^+\mu^-$ system is totally identical to that of the $e^+e^+e^-$ system, differing only by a scaling factor of $m_\mu/m_e$. This is because that in the QED Coulomb potential system, the mass is the only energy scale, with no other dimensional parameters in the potential. As a result, the energy and width are necessarily proportional to the mass, and the rms radii are inversely proportional to the mass.

\begin{table}[htbp] 
\renewcommand{\arraystretch}{1.4}
\centering
\caption{\label{tab:3body_mumumu} The complex energies $\Delta E-i \Gamma / 2$ (in keV), spin configuration and rms radii (in pm) of the $l_1^+l_2^+l_3^-=\mu^+\mu^+\mu^-$ bound and resonant states with $J^P=\frac{1}{2}^-$. $\Delta E$ represents the binding energy relative to the three body threshold. In the ``type" column, ``B" represents bound states, while ``R" denotes resonant states. The spin configuration is represented as $[s_{12},s_{3}]_S$. }
\begin{tabular*}{\hsize}{@{}@{\extracolsep{\fill}}lccccc@{}}

\hline\hline
system & $\Delta E-i \Gamma / 2$ & type & configuration &$r^{\mathrm{rms}}_{l^+l^+}$ &$r^{\mathrm{rms}}_{l^+l^{(')-}}$\\

\hline 
$\mu^+\mu^+\mu^-$ & $-1.47$ & B & $[0,\frac{1}{2}]_{\frac{1}{2}}$ & $2.5$ & $1.8$   \\
& $-0.42$ & R & $[0,\frac{1}{2}]_{\frac{1}{2}}$ & $6.4$ & $4.0$ \\
& $-0.24-0.002i$ & R & $[0,\frac{1}{2}]_{\frac{1}{2}}$ & $8.9$ & $6.3$  \\
& $-0.20$ & R & $[0,\frac{1}{2}]_{\frac{1}{2}}$ & $12.7$ & $7.8$  \\

\hline \hline
\end{tabular*}
\end{table}

\subsubsection{$e^+e^+\mu^-$}\label{subsubsec:eemu}

\begin{figure*}[htbp]
  \centering
  \includegraphics[width=0.8\textwidth]{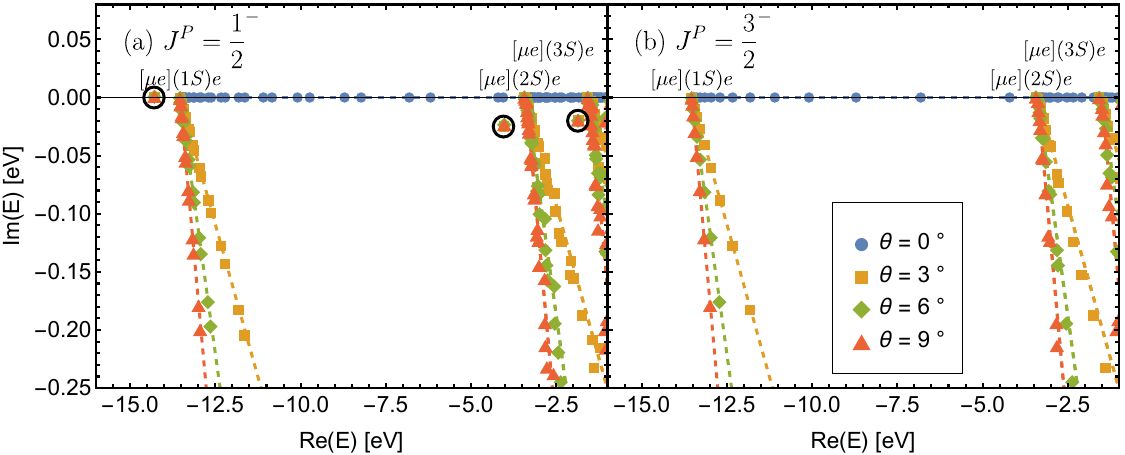} 
  \caption{\label{fig:eemu}The complex energy eigenvalues of the $e^+e^+\mu^-$ states with varying $\theta$ in the CSM. }
    \setlength{\belowdisplayskip}{1pt}
\end{figure*}

The eigenenergies for $e^+e^+\mu^-$ system with $J^{P}=\frac{1}{2}^{-},\frac{3}{2}^{-}$ are shown in Fig.~\ref{fig:eemu}. The bound and resonant states are marked out by black circles. A bound state is found below the lowest $[\mu e](1S)e$ threshold and two resonant states are obtained. Their complex energies, spin configurations, and rms radii are summarized in Table~\ref{tab:3body}. 

Our results for the binding energy of the bound state and the complex energy of the first resonant state under the $2S$ threshold are consistent with those in Ref.~\cite{liverts2013three,frolov2017atomic}. Furthermore, we have identified a new resonant state under the $3S$ threshold.

It remains the case that only the total spin $S=\frac{1}{2}$ system can form bound and resonant states, and their component is also $\left[s_{12}, s_3\right]_S=\left[0, \frac{1}{2}\right]_{\frac{1}{2}}$. The bound state can be treated as the analog of the hydrogen ion. In this system, since there is no annihilation channel, the decay modes of the resonant states correspond to the thresholds below them, i.e. $[\mu e](1 S) e$ for two resonances and $[\mu e](2 S) e$ for the higher resonance.

\subsubsection{$\mu^+\mu^+e^-$}\label{subsubsec:mumue}

\begin{figure*}[htbp]
  \centering
  \includegraphics[width=0.8\textwidth]{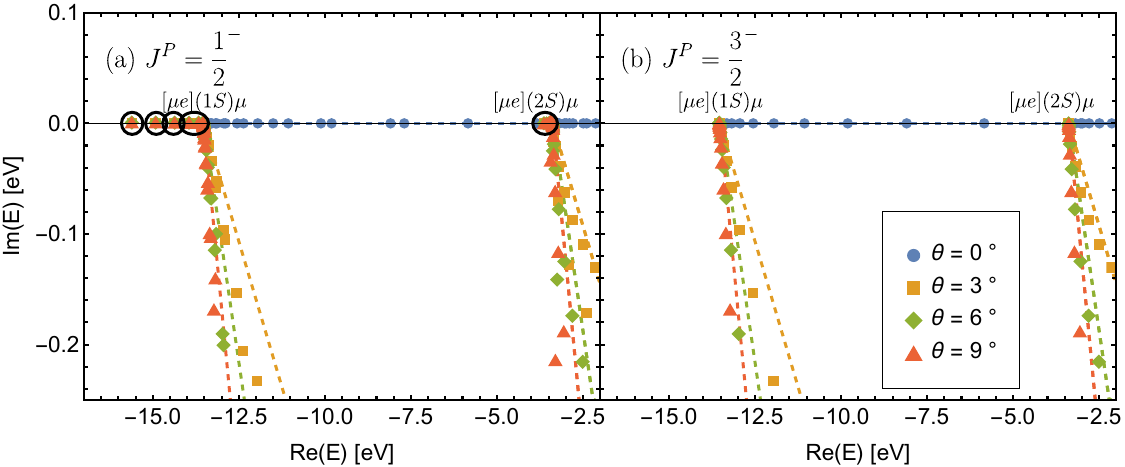} 
  \caption{\label{fig:mumue}The complex energy eigenvalues of the $\mu^+\mu^+e^-$ states with varying $\theta$ in the CSM.}
    \setlength{\belowdisplayskip}{1pt}
\end{figure*}

In the $\mu^+\mu^+e^-$ system, we obtain more bound and resonant states marked out by black circles in Fig.~\ref{fig:mumue}. Their energies, spin configurations, and rms radii are shown in Table~\ref{tab:3body}.

We compare our results with Ref.~\cite{frolov1999bound,liverts2013three} and find that we have identified more bound and resonant states. Although the binding energies of our bound states are not entirely consistent with their results, it is important to note that the focus of this work is not on refining the calculations of bound states but rather on investigating the existence of resonant states. Adjusting the parameters of the Gaussian bases could improve the accuracy of bound state calculations; however, this would compromise the description of the resonant states. Hence, we have opted for a moderate parameter set.

In the $\mu^+\mu^+e^-$ system, the $e^-$ is shared by two $\mu^+$. From the rms radii, it can be seen that $r_{\mu^+\mu^+}$ and $r_{\mu^+e^-}$ are comparable. Although the two $\mu^+$ have large masses and low kinetic energies, which would generally bring them closer, their positive charges result in a repulsive force that causes them to remain far apart. As a result, the distance between the two $\mu^+$ is not significantly smaller than the $\mu^+e^-$ distance.

There is no annihilation channel in this system. So the resonant states can only decay to $[\mu e](1 S) \mu$ through electromagnetic decay. Moreover, we neglect the width of the muon itself.

\subsubsection{$\mu^+e^+\mu^-$ and $\mu^+e^+e^-$}\label{subsubsec:muemu_muee}

We obtain no bound or resonant states in $\mu^+e^+\mu^-$ and $\mu^+e^+e^-$ systems. Our conclusion on the absence of bound states aligns with Ref.~\cite{martin1992stability}. This is a reasonable result, which can be understood through simplified physical pictures. In the $\mu^+e^+\mu^-$ system, the $\mu^+ \mu^-$ pair forms a dipole. Because of their large reduced mass, the $r_{\mu^+ \mu^-}$ is very small, as shown in Table~\ref{tab:2body}, resulting in a small dipole moment and a weak interaction with the extra $e^+$. As a result, they cannot form a three-body bound state. In the $\mu^+e^+e^-$ system, the heavy $\mu^{+}$ can be treated as stationary, with the remaining $e^{-}$ being attractive and the $e^{+}$ repulsive. This makes it difficult to form bound states. In contrast, the $e^{+} e^{+} \mu^{-}$ system can form bound and resonant states because the two $e^{+}$ are both attracted by the $\mu^{-}$, preventing them from flying away.

\subsection{Tetralepton systems}\label{subsec:4l}

\subsubsection{$e^+e^+e^-e^-$ and $\mu^+\mu^+\mu^-\mu^-$}\label{subsubsec:eeee}

\begin{figure*}[htbp]
  \centering
  \includegraphics[width=1.0\textwidth]{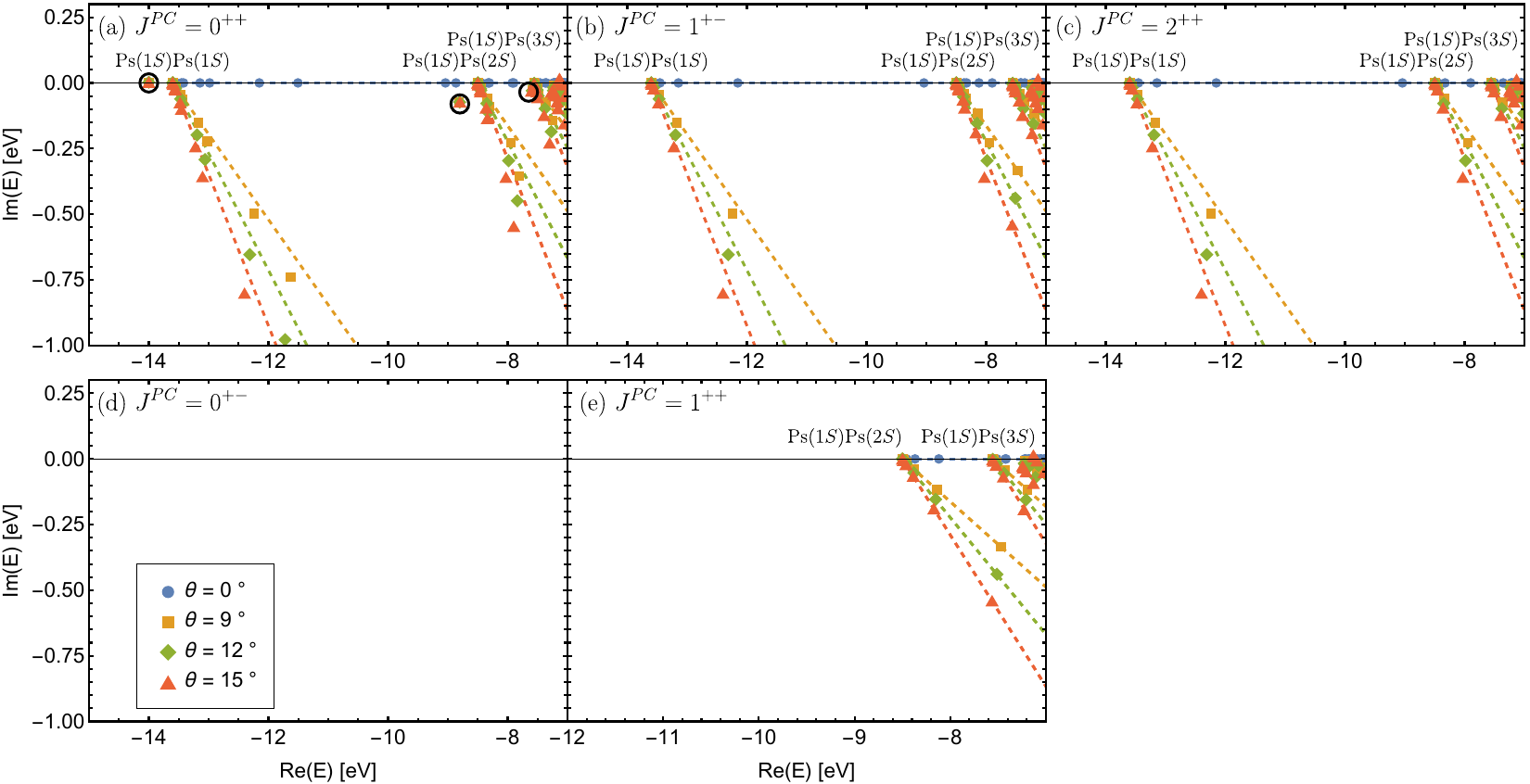} 
  \caption{\label{fig:eeee}The complex energy eigenvalues of the $e^+e^+e^-e^-$ states with varying $\theta$ in the CSM. The solid lines represent the continuum lines rotating along $\mathrm{Arg}(E)=-2\theta$. The resonant states do not shift as $\theta$ changes and are marked out by the black circles. The ``Ps" represents positronium.}
    \setlength{\belowdisplayskip}{1pt}
\end{figure*}

The complex eigenenergies yielded for $e^+e^+e^-e^-$ system with $J^{P}=0^{+\pm},1^{+\pm},2^{++}$ are shown in Fig.~\ref{fig:eeee}. The bound and resonant states are marked out by black circles. We only obtain bound and resonant states in the $J^{P}=0^{++}$ system. A bound state is found below the lowest $\mathrm{Ps}(1S)\mathrm{Ps}(1S)$ threshold and two resonant states are obtained. Their complex energies, spin configurations, and rms radii are summarized in Table~\ref{tab:eeee}.

\begin{table*}[htbp] 
\renewcommand{\arraystretch}{1.4}
\centering
\caption{\label{tab:eeee} The complex energies $\Delta E-i \Gamma / 2$ (in eV), spin configuration and rms radii (in nm) of the $l_1^+l_2^+l_3^-l_4^-=e^+e^+e^-e^-$ bound and resonant states. $\Delta E$ represents the binding energy relative to the four-body threshold. $\Delta E'$ represents the binding energy against dissociation
into two positronium atoms. In the ``type" column, ``B" represents bound states, while ``R" denotes resonant states. The spin configuration is represented as $[s_{12},s_{34}]_S$. In the rightmost column, results from the literature using the variational method with different wave functions are presented for comparison. These include the Hylleraas-type wave function\cite{Ho:1986zz,ho1989resonant} and explicitly correlated Gaussians~\cite{Kinghorn:1993zz,bubin2006nonrelativistic,kozlowski1993nonadiabatic}. All resonance results were obtained using the complex scaling method.
}
\begin{tabular*}{\hsize}{@{}@{\extracolsep{\fill}}lccccccccc@{}}

\hline\hline
$J^{PC}$ & $\Delta E-i \Gamma / 2$ (This work)&$\Delta E'$ (This work) &type & configuration &$r^{\mathrm{rms}}_{e^+e^+}=r^{\mathrm{rms}}_{e^-e^-}$ &$r^{\mathrm{rms}}_{e^+e^-}$  & $\Delta E-i \Gamma  / 2$ & $\Delta E'$ \\

\hline 
$0^{++}$ & $-14.00$ &$-0.40$ & B & $[0,0]_0$ & $0.37$ &$0.29$ &$-14.04$~\cite{Kinghorn:1993zz,bubin2006nonrelativistic} & 
$-0.41$~\cite{Ho:1986zz}, $-0.44$~\cite{kozlowski1993nonadiabatic}\\
& $-8.80-0.07i$  &... & R & $[0,0]_0$ & 0.63  & 0.56 & $-8.52-0.11i$~\cite{ho1989resonant} &... \\
& $-7.59-0.03i$  &...& R & $[0,0]_0$ & 1.30  & 1.16  & $-7.89-0.08i$~\cite{ho1989resonant} &...\\

\hline \hline
\end{tabular*}
\end{table*}

All obtained states have spin configuration $\left[s_{12}, s_{34}\right]_S =[0,0]_0$. This indicates the anti-alignment of spin of the two identical $e^{+}$, and the two identical $e^{-}$ as well. The anti-alignment of spin makes it easier to form bound state because the corresponding spatial wave function is symmetric, allowing it to be an S-wave. The bound and resonant states are only observed in the $0^{++}$ system because it allows for two pairs of identical particles with anti-aligned spins. When the total spin is 1, there can only be one pair of identical particles with anti-aligned spins. When the total spin is 2, both pairs must have aligned spins. The roughly equivalent rms radii of the bound and resonant states show an even distribution of their spatial structure. This is consistent with the spatial structure of the resonant states previously obtained in the fully charm tetraquark system~\cite{Wu:2024hrv}, confirming the similarity between the fully heavy tetraquark system and the tetralepton system.

\begin{figure*}[htbp]
  \centering
  \includegraphics[width=1.0\textwidth]{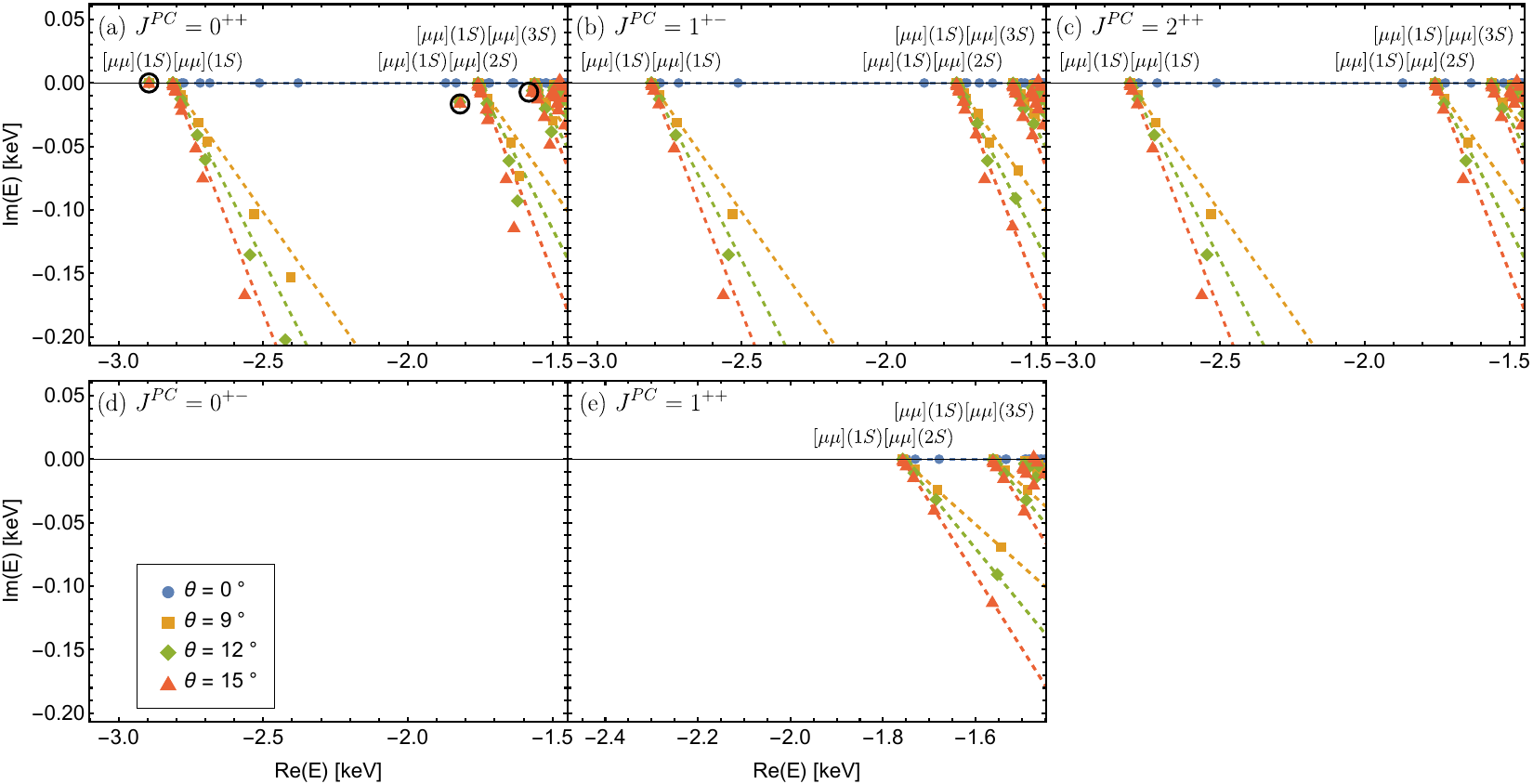} 
  \caption{\label{fig:mumumumu}The complex energy eigenvalues of the $\mu^+\mu^+\mu^-\mu^-$ states with varying $\theta$ in the CSM. The solid lines represent the continuum lines rotating along $\mathrm{Arg}(E)=-2\theta$. The resonant states do not shift as $\theta$ changes and are marked out by the black circles.}
    \setlength{\belowdisplayskip}{1pt}
\end{figure*}

The results of $\mu^+\mu^+\mu^-\mu^-$ system are shown in Fig.~\ref{fig:mumumumu} and Table~\ref{tab:mumumumu}. The pattern of the $\mu^+\mu^+\mu^-\mu^-$ system is identical to that of the $e^+e^+e^-e^-$ system. The reason is the same as for the $\mu^+\mu^+\mu^-$ system: since mass is the only energy scale in the QED Coulomb interaction system, the energy spectra are necessarily proportional to the mass.

\begin{table}[htbp] 
\renewcommand{\arraystretch}{1.4}
\centering
\caption{\label{tab:mumumumu} The complex energies $\Delta E-i \Gamma / 2$ (in keV), spin configuration and rms radii (in pm) of the $l_1^+l_2^+l_3^-l_4^-=\mu^+\mu^+\mu^-\mu^-$ bound and resonant states. $\Delta E$ represents the binding energy relative to the four body threshold. In the ``type" column, ``B" represents bound states, while ``R" denotes resonant states. The spin configuration is represented as $[s_{12},s_{34}]_S$. }
\begin{tabular*}{\hsize}{@{}@{\extracolsep{\fill}}lccccc@{}}

\hline\hline
$J^{PC}$ & $\Delta E-i \Gamma / 2$ &type & configuration &$r^{\mathrm{rms}}_{\mu^+\mu^+}=r^{\mathrm{rms}}_{\mu^-\mu^-}$ &$r^{\mathrm{rms}}_{\mu^+\mu^-}$  \\

\hline 
$0^{++}$ & $-2.89$ & B & $[0,0]_0$ & $1.8$ &$1.4$   \\
& $-1.82-0.015i$  & R & $[0,0]_0$ & 3.1  & 2.7   \\
& $-1.57-0.006i$ & R & $[0,0]_0$ & 6.3  & 5.6 \\

\hline \hline
\end{tabular*}
\end{table}

\subsubsection{$\mu^+\mu^+e^-e^-$}\label{subsubsec:mumuee}

\begin{figure*}[htbp]
  \centering
  \includegraphics[width=1.0\textwidth]{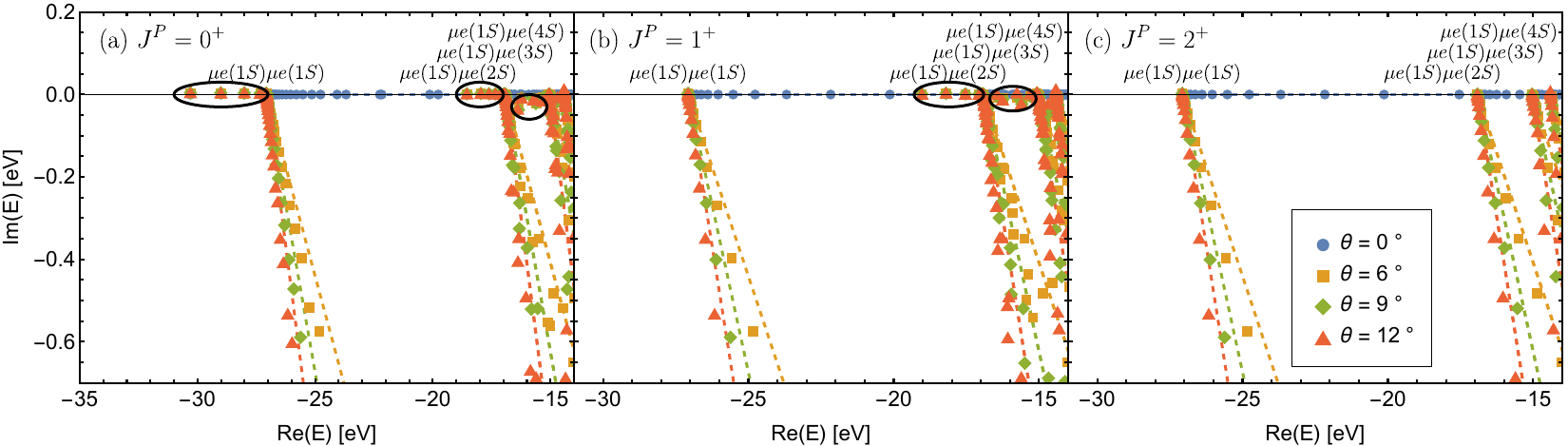} 
  \caption{\label{fig:mumuee}The complex energy eigenvalues of the $\mu^+\mu^+e^-e^-$ states with varying $\theta$ in the CSM. The solid lines represent the continuum lines rotating along $\mathrm{Arg}(E)=-2\theta$. The resonant states do not shift as $\theta$ changes and are marked out by the black circles.}
    \setlength{\belowdisplayskip}{1pt}
\end{figure*}

In the $\mu^+\mu^+e^-e^-$ system, we obtain more bound and resonant states marked out by black circles in Fig.~\ref{fig:mumuee}. Their energies, spin configurations, and rms radii are shown in Table~\ref{tab:mumuee}. The lowest bound state is $-30.30$ eV, which is consistent with the ground-state energy of $-31.05$ eV in Ref.~\cite{rebane2012existence}.

The results show that bound and resonant states exist only in the $0^+$ and $1^+$ channels, with bound states appearing exclusively in the $0^+$ system. The reasons are as follows. In the $0^+$ case, both $\mu^+\mu^+$ and $e^-e^-$ can have their spins anti-aligned, resulting in symmetric spatial wave function, which minimizes the energy and allows the formation of both bound and resonant states. In the $1^+$ case, the total spin must be combined to $1$, so only one pair can have anti-aligned spins (S-wave). This configuration cannot form bound states but can still produce resonant states. Compared to the $e^+e^+e^-e^-$ system with $1^+$, where neither bound nor resonant states can form, the $\mu^+\mu^+$ pair has larger mass and lower kinetic energy, which makes the formation of resonant states possible, though bound states remain unlikely. In the $2^+$ case, neither pair can have anti-aligned spins and no resonant/bound states can form.

Note that in a system with positive parity, it is still possible for the spatial wave function to be exchange antisymmetric only between one pair of identical particles. This component arises from decomposing the S-wave lepton-antilepton pair structure into the Jacobi coordinates of the dilepton-antidilepton structure~\cite{Hiyama:2003cu}. Its parity remains positive because, after decomposition, not only is the spatial wave function between one pair of identical particles in a higher partial wave, but the spatial wave function between the dilepton and antidilepton is also in a higher partial wave, collectively forming a total orbital angular momentum of zero.

For the $\mu^+\mu^+e^-e^-$ bound states, the spatial structure is similar to the covalent bond structure of the hydrogen molecule. A pair of electrons is shared between the two $\mu^+$. The distance $r_{\mu^+e^-}$ is comparable to the two-body $\mu^+e^-$ case. The two heavy $\mu^+$ do not get very close to each other due to their like charges, unlike in the doubly heavy tetraquark systems, where two heavy quarks form an attractive color $\bar{3}_c$ configuration, bringing them closer.

In the $ 1^{+}$ system, the spin components of the resonant states can be either $\left[s_{12}, s_{34}\right]_S=$ $[0,1]_1$ or $[1,0]_1$. For states with leptons $i,j$ forming $s_{i j}=1$, the corresponding $r_{i j}$ is generally larger.

It should be noted that in the energy range of $[-16,-14]\ \mathrm{eV}$, there exists $\left(\mu^{+} \mu^{+} e^{-}\right) e^{-}$ three-body thresholds originating from the bound states in Fig.~\ref{fig:mumue}. However, including the $K$-type Jacobi structure in the calculation would introduce significant complexity, making the extraction of resonant states really difficult. Therefore, $K$-type structures are not included in our calculation. We have checked numerically that the inclusion of $K$-type structures would only affect this specific energy range and would not impact states below $-16\ \mathrm{eV}$.

In the two body Coulomb potential system, the energy of the P-wave state satisfies $E(1 P)=E(2 S)$. Therefore, the absence of the P-wave in this work does not affect the resonant states below the $\mu e(1 S) \mu e(2 S)$ threshold.

\begin{table*}[htbp] 
\renewcommand{\arraystretch}{1.4}
\centering
\caption{\label{tab:mumuee} The complex energies $\Delta E-i \Gamma / 2$ (in eV), spin configuration and rms radii (in nm) of the $l_1^+l_2^+l_3^-l_4^-=\mu^+\mu^+e^-e^-$ bound and resonant states. $\Delta E$ represents the binding energy relative to the four body threshold. In the ``type" column, ``B" represents bound states, while ``R" denotes resonant states. The spin configuration is represented as $[s_{12},s_{34}]_S$. }
\begin{tabular*}{\hsize}{@{}@{\extracolsep{\fill}}lcccccc@{}}

\hline\hline
$J^{P}$ & $\Delta E-i \Gamma / 2$ &type  & configuration &$r^{\mathrm{rms}}_{\mu^+\mu^+}$ &$r^{\mathrm{rms}}_{e^-e^-}$ &$r^{\mathrm{rms}}_{\mu^+e^-}$\\

\hline 
$0^{+}$ & $-30.30$ & B & $[0,0]_0$ & $0.08$ &$0.14$ &$0.10$ \\
& $-29.01$ & B & $[0,0]_0$ & $0.11$ &$0.16$ &$0.11$ \\
& $-28.01$ & B & $[0,0]_0$ & $0.13$ &$0.18$ &$0.13$ \\
& $-27.34$ & B & $[0,0]_0$ & $0.18$ &$0.22$ &$0.16$ \\
& $-18.55$ & R & $[0,0]_0$ & $0.12$ &$0.41$ &$0.29$ \\
& $-17.96$ & R & $[0,0]_0$ & $0.16$ &$0.41$ &$0.29$ \\
& $-17.61$ & R & $[0,0]_0$ & $0.21$ &$0.41$ &$0.30$ \\
& $-17.34$ & R & $[0,0]_0$ & $0.26$ &$0.44$ &$0.31$ \\
& $-17.12$ & R & $[0,0]_0$ & $0.32$ &$0.48$ &$0.34$ \\
& $-16.98$ & R & $[0,0]_0$ & $0.41$ &$0.56$ &$0.40$ \\

& $-16.33-0.03i$ & R & $[0,0]_0$ & $0.12$ &$1.27$ &$0.90$ \\
& $-16.22-0.01i$ & R & $[0,0]_0$ & $0.15$ &$0.91$ &$0.65$ \\
& $-15.72-0.01i$ & R & $[0,0]_0$ & $0.18$ &$0.86$ &$0.61$ \\
& $-15.60-0.02i$ & R & $[0,0]_0$ & $0.16$ &$1.30$ &$0.92$ \\
& $-15.33-0.01i$ & R & $[0,0]_0$ & $0.24$ &$0.85$ &$0.60$ \\

\hline 
$1^{+}$ & $-18.20$ & R & $[0,1]_1$ & $0.13$ &$0.39$ &$0.27$ \\
& $-17.53$ & R & $[0,1]_1$ & $0.16$ &$0.41$ &$0.29$ \\
& $-17.07$ & R & $[1,0]_1$ & $0.35$ &$0.48$ &$0.34$ \\
& $-17.04$ & R & $[0,1]_1$ & $0.16$ &$0.66$ &$0.47$ \\
& $-17.03$ & R & $[0,1]_1$ & $0.17$ &$0.64$ &$0.45$ \\
& $-16.95$ & R & $[1,0]_1$ & $0.46$ &$0.58$ &$0.41$ \\

& $-16.42-0.01i$ & R & $[0,1]_1$ & $0.12$ &$1.22$ &$0.87$ \\
& $-16.95-0.01i$ & R & $[0,1]_1$ & $0.14$ &$0.95$ &$0.68$ \\
& $-15.75-0.01i$ & R & $[0,1]_1$ & $0.17$ &$1.00$ &$0.72$ \\
& $-15.58-0.02i$ & R & $[0,1]_1$ & $0.17$ &$1.18$ &$0.84$ \\
& $-15.30$ & R & $[0,1]_1$ & $0.23$ &$0.88$ &$0.63$ \\

\hline \hline
\end{tabular*}
\end{table*}

\subsubsection{$\mu^+e^+\mu^-e^-$}\label{subsubsec:muemue}

We find no bound or resonant states in the $\mu^+e^+\mu^-e^-$ system, consistent with the conclusions in Ref.~\cite{Bressanini:1997zz,rebane2003binding,gridnev2005proof}. The reason is the same as the $\mu^+e^+\mu^-$ system. The heavy $\mu^+\mu^-$ pair forms a dipole with small dipole moment, resulting in a weak interaction with the remaining $e^+e^-$. 

\section{Summary and Discussion}~\label{sec:sum}

We calculate the mass spectrum of the S-wave trilepton systems $e^+e^+e^-$, $\mu^+\mu^+\mu^-$, $e^+e^+\mu^-$, $\mu^+\mu^+e^-$, $e^+\mu^+e^-$, $\mu^+e^+\mu^-$ and tetralepton systems $e^+e^+e^-e^-$, $\mu^+\mu^+\mu^-\mu^-$, $\mu^+\mu^+e^-e^-$, $\mu^+e^+\mu^-e^-$ using the QED Coulomb potential. We employ the GEM to solve the $n$-body Schr\"odinger equation, and the CSM to identify genuine resonant states from scattering states.

We obtain a series of bound and resonant states in the trilepton systems $e^+e^+e^-$, $\mu^+\mu^+\mu^-$, $e^+e^+\mu^-$, $\mu^+\mu^+e^-$, as well as tetralepton systems $e^+e^+e^-e^-$, $\mu^+\mu^+\mu^-\mu^-$, $\mu^+\mu^+e^-e^-$. Their energies range from -30 eV to -1 eV relative to the total mass of three or four leptons, with their widths spanning from less than 0.01 eV to approximately 0.07 eV. We find no bound or resonant states in the trilepton $e^+\mu^+e^-$, $\mu^+e^+\mu^-$ systems and tetralepton $\mu^+e^+\mu^-e^-$ system.

The spin configurations of the majority of bound and resonant states are such that the spins of identical particles are anti-aligned. Except for the resonant states in the $ \mu^+ \mu^+ e^- e^-$ system with $1^+$, where one pair of identical particles is anti-aligned while the other pair is aligned. Based on the results of the rms radii, the states in the $e^+e^+e^-$, $\mu^+\mu^+\mu^-$, $\mu^+\mu^+e^-$, and $\mu^+\mu^+e^-e^-$ systems exhibit a covalent bond-like spatial structure, while in the $e^+e^+e^-e^-$ and $\mu^+\mu^+\mu^-\mu^-$ states, the distributions are even.

Comparing tetralepton and tetraquark systems can provide us with more insights. A comparison between the $e^+e^+e^-e^-$ (or $\mu^+\mu^+\mu^-\mu^-$) system and the fully heavy tetraquark system~\cite{Wu:2024euj,Wu:2024hrv} reveals that all the states exhibit a uniform distribution, suggesting a similarity between these two types of systems. Resonant states above the $M(1S)M'(2S)$ threshold are obtained in both systems.  However, in the fully heavy tetraquark system, no bound states or resonant states are obtained below the $M(1 S) M^{\prime}(2 S)$ dimeson thresholds. Moreover, in the fully heavy tetraquark system, resonant states can form for all $J^{P C}$, whereas in the $e^{+} e^{+} e^{-} e^{-}$ system, only the $0^{++}$ system can form bound or resonant states. The difference may arise from the fact that, although the fully heavy tetraquark system is also dominated by the Coulomb potential, quarks have an additional color degree of freedom compared to leptons, making the system more complex. Moreover, the Coulomb term in the interaction between quarks also includes a color-dependent $\lambda_i \cdot \lambda_j$ matrix element, which means that the attraction and repulsion between quarks are not as straightforward as in the charged lepton systems.

This comparison highlights that the additional color degree of freedom in QCD systems, compared to QED systems, leads to the disappearance of lower-energy bound states and resonant states. However, it also enables a greater variety of $J^{PC}$ quantum numbers to produce resonant states.

The presence of multi-lepton resonant states themselves holds significant importance. This work, as the first to predict tetralepton resonant states beyond the simplest 4-body system $\mathrm{Ps}_2$, marks the advent of a new class of exotic states. These tetralepton states, once beyond experimental reach, now stand on the cusp of possible realization with the rapid advancements in experimental technologies. The Super $\tau$-Charm Facility\cite{Achasov:2023gey} in the future, holds great promise for the production and study of these intriguing states. Furthermore, with improvements in muon beam technology, there exists the potential to generate these trilepton and tetralepton states, paving the way for new insights into multi-lepton systems and expanding our understanding of exotic states in particle physics.

\begin{appendix}

\section{Gaussian bases parameters}~\label{app:Gaussian_par}

The settings of the basis parameters for each system in our calculation are:
\begin{itemize}
\setlength{\itemindent}{-16pt}
\item $e^+e^+e^-$:
\begin{multline}
\left\{
\begin{array}{ll}
e^{+}-e^{+}: & r_n \in [0.02, 3.0]\ \mathrm{nm},\ n=12\   \\
(e^+ e^+)-e^-: & r_n \in [0.05, 5.2]\ \mathrm{nm} ,\ n=12\  \\
e^{+}-e^{-}:&  r_n \in [0.02, 3.0]\ \mathrm{nm} ,\ n=20\  \\
(e^{+} e^{-})-e^{+}: & r_n \in [0.05, 5.0]\ \mathrm{nm} ,\ n=20\ \\
\end{array}
\right.
\end{multline}
\end{itemize}

\begin{itemize}
\setlength{\itemindent}{-16pt}
\item $\mu^+\mu^+\mu^-$:
\begin{multline}
\left\{
\begin{array}{ll}
\mu^{+}-\mu^{+}: & r_n \in [0.097, 14.5]\ \mathrm{pm},\ n=12\   \\
(\mu^+ \mu^+)-\mu^-: & r_n \in [0.24, 25.1]\ \mathrm{pm} ,\ n=12\  \\
\mu^{+}-\mu^{-}:&  r_n \in [0.097, 14.5]\ \mathrm{pm} ,\ n=20\  \\
(\mu^{+} \mu^{-})-\mu^{+}: & r_n \in [0.24, 24.2]\ \mathrm{pm} ,\ n=20\ \\
\end{array}
\right.
\end{multline}
\end{itemize}

\begin{itemize}
\setlength{\itemindent}{-16pt}
\item $e^+e^+\mu^-$:
\begin{multline}
\left\{
\begin{array}{ll}
e^{+}-e^{+}: & r_n \in [0.01, 1.06]\ \mathrm{nm},\ n=12\   \\
(e^+ e^+)-\mu^-: & r_n \in [0.03, 3.0]\ \mathrm{nm} ,\ n=12\  \\
e^{+}-\mu^{-}:&  r_n \in [0.01, 1.7]\ \mathrm{nm} ,\ n=30\  \\
(e^{+} \mu^{-})-e^{+}: & r_n \in [0.03, 3.0]\ \mathrm{nm} ,\ n=30\ \\
\end{array}
\right.
\end{multline}
\end{itemize}

\begin{itemize}
\setlength{\itemindent}{-16pt}
\item $\mu^+\mu^+e^-$:
\begin{multline}
\left\{
\begin{array}{ll}
\mu^{+}-\mu^{+}: & r_n \in [0.0001, 0.01]\ \mathrm{nm},\ n=12\   \\
(\mu^+ \mu^+)-e^-: & r_n \in [0.015, 3.0]\ \mathrm{nm} ,\ n=12\  \\
\mu^{+}-e^{-}:&  r_n \in [0.01, 1.7]\ \mathrm{nm} ,\ n=30\  \\
(\mu^{+} e^{-})-\mu^{+}: & r_n \in [0.015, 1.5]\ \mathrm{nm} ,\ n=30\ \\
\end{array}
\right.
\end{multline}
\end{itemize}

\begin{itemize}
\setlength{\itemindent}{-16pt}
\item $\mu^+e^+e^-$:
\begin{multline}
\left\{
\begin{array}{ll}
\mu^{+}-e^{+}: & r_n \in [0.01, 1.7]\ \mathrm{nm},\ n=12\   \\
(\mu^+ e^+)-e^-: & r_n \in [0.02, 3.0]\ \mathrm{nm} ,\ n=12\  \\
\mu^{+}-e^{-}:&  r_n \in [0.01, 1.7]\ \mathrm{nm} ,\ n=30\  \\
(\mu^{+} e^{-})-e^{+}: & r_n \in [0.03, 3.0]\ \mathrm{nm} ,\ n=30\ \\
e^{+}-e^{-}:&  r_n \in [0.01, 1.7]\ \mathrm{nm} ,\ n=30\  \\
(e^{+} e^{-})-\mu^{+}: & r_n \in [0.03, 3.0]\ \mathrm{nm} ,\ n=30\ \\
\end{array}
\right.
\end{multline}
\end{itemize}

\begin{itemize}
\setlength{\itemindent}{-16pt}
\item $\mu^+e^+\mu^-$:
\begin{multline}
\left\{
\begin{array}{ll}
\mu^{+}-e^{+}: & r_n \in [0.01, 1.7]\ \mathrm{nm},\ n=12\   \\
(\mu^+ e^+)-\mu^-: & r_n \in [2.0, 10.5]\ \mathrm{nm} ,\ n=12\  \\
\mu^{+}-\mu^{-}:&  r_n \in [0.00003, 0.012]\ \mathrm{nm} ,\ n=30\  \\
(\mu^{+} \mu^{-})-e^{+}: & r_n \in [0.01, 5.0]\ \mathrm{nm} ,\ n=30\ \\
e^{+}-\mu^{-}:&  r_n \in [0.01, 1.7]\ \mathrm{nm} ,\ n=30\  \\
(e^{+} \mu^{-})-\mu^{+}: & r_n \in [2.0, 10.5]\ \mathrm{nm} ,\ n=30\ \\
\end{array}
\right.
\end{multline}
\end{itemize}

\begin{itemize}
\setlength{\itemindent}{-16pt}
\item $e^+e^+e^-e^-$:
\begin{multline}
\left\{
\begin{array}{ll}
e^{+}-e^{+} \text { or } e^{-}-e^{-}: & r_n \in [0.02, 3.0]\ \mathrm{nm},\ n=12\   \\
(e^+ e^+)-(e^- e^-): & r_n \in [0.02, 3.0]\ \mathrm{nm},\ n=12\  \\
e^{+}-e^{-}: & r_n \in [0.02, 3.0]\ \mathrm{nm},\ n=12\  \\
(e^{+} e^{-})-(e^{+} e^{-}): & r_n \in [0.02, 6.0]\ \mathrm{nm},\ n=12\ \\
\end{array}
\right.
\end{multline}
\end{itemize}

\begin{itemize}
\setlength{\itemindent}{-16pt}
\item $\mu^+\mu^+\mu^-\mu^-$:
\begin{multline}
\left\{
\begin{array}{ll}
\mu^{+}-\mu^{+} \text { or } \mu^{-}-\mu^{-}: & r_n \in [0.097, 14.5]\ \mathrm{pm},\ n=12\   \\
(\mu^+ \mu^+)-(\mu^- \mu^-): & r_n \in [0.097, 14.5]\ \mathrm{pm},\ n=12\  \\
\mu^{+}-\mu^{-}: & r_n \in [0.097, 14.5]\ \mathrm{pm},\ n=12\  \\
(\mu^{+} \mu^{-})-(\mu^{+} \mu^{-}): & r_n \in [0.097, 29.0]\ \mathrm{pm},\ n=12\ \\
\end{array}
\right.
\end{multline}
\end{itemize}

\begin{itemize}
\setlength{\itemindent}{-16pt}
\item $\mu^+\mu^+e^-e^-$:
\begin{multline}
\left\{
\begin{array}{ll}
\mu^{+}-\mu^{+}: & r_n \in [0.0001, 0.01]\ \mathrm{nm},\ n=12\   \\
e^{-}-e^{-}: & r_n \in [0.01, 1.06]\ \mathrm{nm},\ n=12\   \\
(\mu^{+}\mu^{+})-(e^- e^-): & r_n \in [0.025, 0.7]\ \mathrm{nm},\ n=12\  \\
\mu^{+}-e^{-}: & r_n \in [0.01, 1.06]\ \mathrm{nm},\ n=12\  \\
(\mu^{+} e^{-})-(\mu^{+} e^{-}): & r_n \in [0.025, 0.7]\ \mathrm{nm},\ n=30\  \\
\end{array}
\right.
\end{multline}
\end{itemize}

\begin{itemize}
\setlength{\itemindent}{-16pt}
\item $\mu^+e^+\mu^-e^-$:
\begin{multline}
\left\{
\begin{array}{ll}
\mu^{+}-e^{+} \text { or } \mu^{-}-e^{-}: & r_n \in [1.0, 5.0]\ \mathrm{nm},\ n=12\   \\
(\mu^+ e^+)-(\mu^- e^-): & r_n \in [3.5, 4.0]\ \mathrm{nm},\ n=12\  \\
\mu^{+}-\mu^{-}: & r_n \in [0.00005, 0.0086]\ \mathrm{nm},\ n=12\  \\
e^{+}-e^{-}: & r_n \in [0.05, 3.5]\ \mathrm{nm},\ n=12\  \\
(\mu^{+} \mu^{-})-(e^{+} e^{-}): & r_n \in [0.025, 0.7]\ \mathrm{nm},\ n=12\ \\
\mu^{+}-e^{-} \text { or } e^{+}-\mu^{-}: & r_n \in [0.01, 1.06]\ \mathrm{nm},\ n=12\  \\
(\mu^{+} e^{-})-(e^{+} \mu^{-}): & r_n \in [3.45, 4.0]\ \mathrm{nm},\ n=12\ \\
\end{array}
\right.
\end{multline}
\end{itemize}

In some cases, we employ a larger number of basis functions to better describe the continuum states.

\end{appendix}

\begin{acknowledgements}

We are grateful to Wei-Lin Wu, Yan-Ke Chen, Zi-Yang Lin, Hui-Min Yang and Jun-Zhang Wang for the helpful discussions. 
This project was supported by the National Natural Science Foundation of China (Grant No. 12475137), and ERC NuclearTheory (Grant No. 885150). The computational resources were supported by High-performance Computing Platform of Peking University.

\end{acknowledgements}

\section*{DATA AVAILABILITY STATEMENT}

The data supporting the findings of this study is available on OSF repository~\cite{ma_2025_QED_data} or can be obtained directly from the authors upon reasonable request.

\bibliography{Ref}

\end{document}